# Fluid-structure interaction of a large ice sheet in waves

Luofeng Huang[a, *], Kang Ren[a], Minghao Li[b], Željko Tuković[c], Philip Cardiff[d], Giles Thomas[a]

[a]Department of Mechanical Engineering, University College London, United Kingdom

[b]Department of Mechanics and Maritime Science, Chalmers University of Technology, Sweden

[c]Faculty of Mechanical Engineering and Naval Architecture, University of Zagreb, Croatia

[d]School of Mechanical and Materials Engineering, University College Dublin, Ireland

**Abstract**

With global warming, the ice-covered areas in the Arctic are being transformed into open water. This provides increased impetus for extensive maritime activities and attracts research interests in sea ice modelling. In the polar region, ice sheets can be several kilometres long and subjected to the effects of ocean waves. As its thickness to length ratio is very small, the wave response of such a large ice sheet, known as its hydroelastic response, is dominated by an elastic deformation rather than rigid body motions. In the past 25 years, sea ice hydroelasticity has been widely studied by theoretical models; however, recent experiments indicate that the ideal assumptions used for these theoretical models can cause considerable inaccuracies. This work proposes a numerical approach based on OpenFOAM to simulate the hydroelastic wave-ice interaction, with the Navier-Stokes equations describing the fluid domain, the St. Venant Kirchhoff solid model governing the ice deformation and a coupling scheme to achieve the fluid-structure interaction. Following validation against experiments, the proposed model has been shown capable of capturing phenomena that have not been included in current theoretical models. In particular, the developed model shows the capability to predict overwash, which is a ubiquitous polar phenomenon reported to be a key gap. The present model has the potential to be used to study wave-ice behaviours and the coupled wave-ice effect on marine structures.

*Corresponding author: ucemlhu@ucl.ac.uk

## 1. Introduction

Since the late 1970s when satellite observations began, the Arctic ice extent has decreased rapidly with the effect of global warming. The summer ice coverage is reducing at a rate of 12.9 ± 1.47% per decade (Stroeve et al., 2012), evidenced by an obvious transition from ice-covered areas to open water. Multiple predictive models have indicated that the retreat of the Arctic ice will not slow down (Wadhams, 2017); the Arctic is evolving into a navigable ocean (Smith and Stephenson, 2013) and vast quantities of natural resources such as oil, gas and minerals are becoming extractible. This suggests that large-scale activities of Arctic shipping and resource exploitation may occur very soon, attracting special research interest and significant investment from commercial stakeholders.

The existence of sea ice distinguishes the polar condition from that of an open ocean, which influences potential maritime activities and provides challenges for the safety and optimisation of polar engineering. Sea ice can directly collide with marine structures (McGovern and Bai, 2014a; Guo et al., 2018; Luo et al., 2018) and induce wave scattering/attenuation (Bennetts and Williams, 2015; Toffoli et al., 2015; Montiel et al., 2018) thus indirectly varying the hydrodynamic response of passing vessels (Ren et al., 2016) and offshore platforms (Ren et al., 2018). To date, although the ultimate goal of Arctic engineering is to predict the coupling wave-ice-structure mechanism of the above processes, research has focussed on wave-ice interactions (Squire et al., 1995; Squire, 2007). Through the development of accurate models for wave-ice behaviours, appropriate approaches will be established towards the wave-ice-structure coupling.

In the Arctic, sea ice typically exists as level ice sheets that can be kilometres long, or broken into small pancake-shaped ice floes floating on the sea surface (Thomson et al., 2018). For the ice floe, according to its small dimension compared with the dominant wavelength, it usually undergoes rigid-body motions induced by the ocean waves. A good description of such a behaviour may be found in a handful of publications (McGovern and Bai, 2014b; Meylan et al., 2015; Yiew et al., 2016, 2017). However, for a large ice sheet, as its thickness is very small compared to the length, it may exhibit localised vibrations under a continuous wave excitation. In such a situation, the wave response of the ice sheet is dominated by the elastic deformations other than rigid body motions, known as the hydroelasticity of sea ice. A review of this phenomenon has been given by Squire (2011), where the author notes its prediction is a key challenge of polar engineering. Noting that the sea ice retreat is a process of volume loss, apart from the decline of extent, the Arctic ice thickness has reduced by more than 50% in the past 50 years (Laxon et al., 2013). The effect of sea ice hydroelasticity on these increasingly thin sheets has become significant.

Studies on sea ice hydroelasticity have mainly addressed the wave-induced ice deformation and the transmission/reflection of surface waves encountering an ice sheet. In most cases, the ice sheet is modelled as a thin elastic plate subjected to regular ocean waves. Experimentally, Meylan et al. (2015)

conducted wave basin tests to measure the wave response of a plastic plate. They found that the plate bends due to wave propagation. Performing similar experiments, Sree et al. (2018) observed that high aspect-ratio plates tended to follow the wave-shape, and reported wave attenuation due to the viscous effect between water and the plate bottom. Sree et al. (2017) also found the wavelength and wave celerity inside the plate are larger than those of an open water situation. Dolatshah et al. (2018) conducted real ice tests to study the wave-induced ice vibration and breakup. When encountering an ice sheet, waves partially pass through and are partially reflected, for which Nelli et al. (2017) measured the proportion of the transmitted/reflected waves, expressed as transmission/reflection coefficients. With increasing wavelength, waves were found to be transmitted more and reflected less. Their experiments also indicated the reflection coefficient is insensitive to wave amplitude, while the transmission coefficient can be reduced by the energy dissipative "overwash" phenomenon. Overwash is defined as waves running over the plate surface and is strongly dictated by the incoming wave amplitude. The depths of overwash water at different wave conditions were reported by Skene et al. (2015).

Due to the prohibitive cost of experimental testing, theoretical models have the potential to provide more efficient and economical solutions. Theoretical predictions started with obtaining the transmission and reflection coefficients of surface waves propagating against a semi-infinite ice sheet. These models were based on linearised theories that can also be applied to Very Large Floating Structures (VLFS) (Squire, 2008). In this approach, potential flow theory is employed in the fluid domain and the ice sheet is treated as a linear elastic thin plate. Fox and Squire (1994) considered the problem of wave transmission/reflection from open water into an ice sheet. The Eigenfunction Expansion Method (EEM) was adopted for the velocity potentials underneath the open water surface and ice sheet, and an iterative conjugate gradient method was used to impose continuity between these two parts. The transmission and reflection coefficients of waves were obtained, alongside their relationship with the incident wavelength, ice thickness and water depth. Other methods were also well applied to the same case, e.g. Chung and Fox (2002) employed the Wiener-Hopf method; Hermans (2007) used the Green's function method. Although the above studies ignored the submergence of the ice sheet, the ice draught was afterwards included by Bennetts et al. (2007), William and Squire (2008) and William and Porter (2009). Apart from a semi-infinite ice sheet, relevant linear models were also applied to the case of a finite ice sheet. Meylan and Squire investigated the hydroelasticity of a solitary ice sheet (Meylan and Squire, 1993, 1996) and a pair of ice sheets (Meylan and Squire, 1994). Wang and Meylan (2004) used the Green's function method to solve the fluid domain surrounding an ice sheet and calculated the wave-induced ice deformation by the finite-element method. In addition, Smith and Meylan (2011) investigated the influence of ice thickness on wave transmission. More examples can be found in the reviews of Squire et al. (1995, 2007).

Works based on theoretical models have provided great insight into sea ice hydroelasticity; however, they are built upon certain ideal assumptions. For example, they usually ignore the nonlinearity, viscosity and turbulence of the fluid and assume the wave amplitude to be very small. These assumptions exclude some important phenomena, e.g. overwash, which is a highly frequent phenomenon due to the very small freeboard of sea ice. Recent experiments assessed the accuracy of existing theoretical models and demonstrated in certain scenarios these assumptions can cause considerable deviations (Bennetts and Williams, 2015; Yiew et al., 2016). Toffoli et al. (2015) and Nelli et al. (2017) demonstrated the theoretical approach cannot accurately predict the transmission and reflection coefficients when overwash occurs. To obtain a more realistic solution, one approach is to use the Computational Fluid Dynamics (CFD) technique to numerically solve the nonlinear Navier-Stokes equations. The CFD approach allows a fully-matched solution to be achieved between the wave and ice (Huang and Thomas, 2018), while also allowing the inclusion of complex geometry, important for marine structures.

CFD has been widely applied to hydrodynamic problems (Jasak, 2017), e.g. using the open-source code, OpenFOAM (Jasak et al., 2007). Validation against experiments has shown OpenFOAM is capable of accurately simulating regular/irregular wave fields and obtain the static/dynamic response of fixed/floating structures (Jacobsen et al., 2012; Higuera et al., 2013; Chen et al., 2014; Bruinsma et al., 2018). Regarding the problem of wave interaction with a floating ice sheet, Bai et al. (2017) used OpenFOAM to study the wave-induced movement of a small floating plate, which achieved good agreement with the corresponding experiments. However, they assumed the plate to be rigid, which limits this model to be only applicable to cases where the ice dimension is much smaller than the dominant wavelength. For a large ice sheet where the hydroelastic response is significant, the rigid assumption becomes unrealistic. In such a situation, a Fluid-Structure Interaction (FSI) approach is required to obtain the structural solution of ice deformation and couple it with the solution of surrounding fluid domain, i.e. fully simulating the hydroelastic wave-ice interaction.

Tukovic et al. (2007, 2014) developed an FSI code based on OpenFOAM (fsiFoam solver). It employed a partitioned FSI scheme to include the two-way coupling between the fluid and structure, where the fluid and solid solutions are solved separately and coupled via the fluid-solid interface. An advantage of this approach is that it employs the finite-volume method (FVM) for both fluid and solid domains (Tuković et al., 2013; Tukovic and Jasak, 2007). Most current FSI works involve a combination of solvers, usually with a finite-volume solver for the fluid flow and a finite-element solver for the structural analysis, e.g. (McVicar et al., 2018), which requires a third code for coupling, data interpolation and simulation management. Thus, the combined alternative approaches for the fluid and solid domains will tend to increase computational costs and imposes limitations on the coupling method. In contrast, the entirely FVM approach of Tukovic et al. (2014) makes an all-in-one solver under the framework of OpenFOAM. Furthermore, a benefit of its open-source nature is that others can add extended models, e.g. viscoelastic, thermoelastic, and poroelastic solids (Tang et al., 2015; Cardiff et al., 2018).

One gap of the current OpenFOAM FSI approach is that it has only been applied to single-phase fluid modelling (Rege and Hjertager, 2017). Therefore, it has not been applied to ocean engineering applications containing both air and water. In order to simulate hydroelastic problems within OpenFOAM, Huang et al. (2018a, 2018b) incorporated a Volume of Fluid (VOF) (Hirt and Nichols, 1981) approach to model multiphase flows. In this way, the simulation of the hydroelastic interaction between waves and a large ice sheet is possible, and its realisation is presented as follows. This is also the first application of the FSI procedure of Tukovic et al. (2014) to ocean engineering.

This paper starts by introducing the numerical theories and practicalities of building a CFD model to simulate the hydroelastic wave-ice interaction. Subsequently, validation of the model is presented against the experiments of Sree et al. (2017), Nelli et al. (2017) and Skene et al. (2015). Following investigations focus on the wave transmission/reflection induced by sea ice, as this indicates the influence of sea ice on the surrounding wave field, which is a key factor in the hydrodynamics of polar regions. The computational results are also compared with the corresponding analytical solutions predicted by a standard theoretical method, where the inaccuracies existing in current theoretical models are shown to be effectively remedied by the new approach.

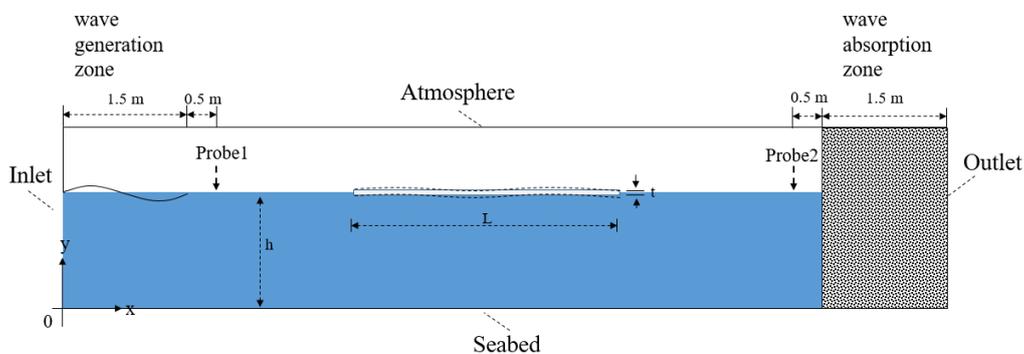

Figure 1: Schematic of the case: a thin ice sheet is floating on the water surface and subjected to incoming waves, with its elastic deformation induced.

## 2. Numerical approach

In contrast to the traditional CFD approach that only solves governing equations for the fluid domain, the FSI approach requires the solutions of both the fluid and solid domains, alongside a coupling scheme to link the solutions together. To achieve this, the computational domain is divided into two parts, namely the fluid sub-domain and the solid sub-domain. A numerical wave tank is established in the fluid sub-domain to generate a regular wave field, and the solid sub-domain is a thin ice sheet floating on the water surface. The case setup and solution process are presented as follows.

2.1 Computational domain and boundary conditions

As shown in Figure 1, a two-dimensional rectangular computational domain was assumed, defined by the Cartesian *xy* coordinate system (indicated by 0 in Figure 1). The *x*-axis is parallel to the undisturbed water surface, and the *y*-axis is positive upwards. The computational domain (12 m long and 1.5 m high) is filled with fresh water to a depth of *h* with air filling the remainder of the fluid sub-domain. At the top boundary of the domain, a static pressure boundary condition is applied to represent atmospheric conditions. The bottom boundary is defined as a no-slip wall to account for the presence of the seabed. The solid sub-domain represents an ice sheet floating on the water surface according to its buoyancy-gravity equilibrium position. The length and thickness of ice are denoted as *L* and *t* respectively.

Periodic regular waves were generated at the inlet boundary (left boundary of the domain), propagating in the positive *x*-direction, and a wave absorption zone was placed by the outlet boundary at the right of the domain to minimise reflection of waves. The wave generation and absorption were realised using the relaxation zone method implemented within the open-source wave toolbox, waves2Foam (Jacobsen et al., 2012). In the relaxation zone method, a spatial weighting factor $\chi$ is introduced. Along the positive direction of x-axis, $\chi$ increases from 0 to 1 in the wave generation zone and decreases from 1 to 0 in the wave absorption zone. A local field $\varphi$ is dependent on $\chi$ as:

$$\phi = \chi \phi_{\text{computed}} + (1-\chi)\phi_{\text{target}} \tag{1}$$

Thus, target results can be obtained at the inlet and outlet boundaries, where $\chi = 0$, while fully computed results can be obtained between the two zones, where $\chi = 1$.

In the wave generation zone, the $\phi_{\text{target}}$ was defined as a regular wave field, according to the linear Stokes's wave theory (Dean and Dalrymple, 1991),

$$\eta = h + \frac{H}{2}\cos(kx - \omega t) \tag{2}$$

$$u = \frac{\pi H}{\delta}\frac{\cosh k(y+h)}{\sinh kh}\cos(kx - \omega t) \tag{3}$$

$$v = \frac{\pi H}{\delta} \frac{sinh k(y+h)}{sinh\, kh} sin(kx - \omega t) \qquad (4)$$

In which $\eta$ is the free surface elevation, $H$ is the wave height (double of the wave amplitude $a$), $\delta$ is the wave period, $k$ is the wave number and $\omega$ is the angular frequency. For the wave properties, $H$ and $\delta$ are given in advance, and the wavelength ($\lambda = 2\pi/k$) was solved by the dispersion relation:

$$k \tanh kh = \kappa, \qquad \text{where} \quad \kappa = \omega^2/g \qquad (5)$$

where g is gravitational acceleration set as 9.81 m/s². In the wave damping zone, the $\phi_{target}$ was set as still water, i.e. $\eta = h$ and $u = v = 0$. Two probes were positioned at upstream and downstream locations to record the time-varying free surface elevation.

2.2 Computational method

The Finite Volume Method (Versteeg and Malalasekera, 2007) was applied to obtain the fluid and structural solutions over a certain time duration. The process includes two types of discretisation, in space and time respectively. In space, the computational domain is divided into a set of non-overlapping hexahedral cells, known as a mesh; in time, the temporal dimension is split into a finite number of timesteps. As this is an FSI case, the computational mesh was divided into two parts, namely a fluid mesh for the fluid sub-domain and a solid mesh for the solid sub-domain, as shown in Figure 2. They are connected by placing their interface boundaries at the same location, where the fluid and solid interface meshes need not be conformal. The fluid mesh is graded towards the free surface area, while the solid mesh density is uniform. The cell numbers of both meshes were determined by sensitivity tests, as presented in Section 3.1. The size of each timestep was determined by a prescribed Courant number (Co) value, according to the expression:

$$Co = \frac{u \Delta t}{\Delta x} < 1 \qquad (6)$$

where $\Delta t$ is the timestep size, $u/\Delta x$ is its normal velocity divided by the distance between the cell centre and the centre of the neighbour cell. Hereby, $\Delta t$ in this study was given at 0.0005 s.

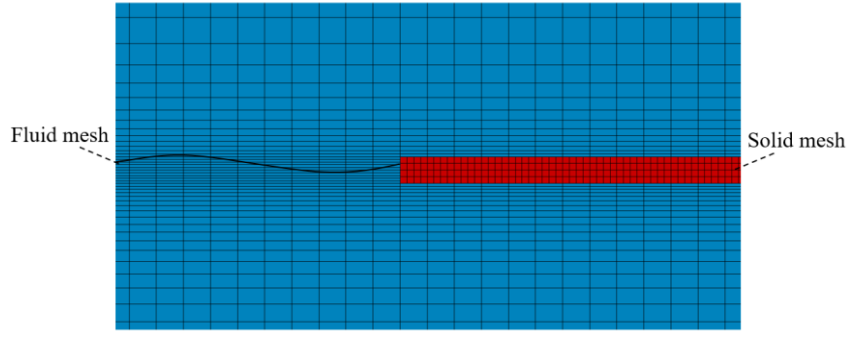

Figure 2: Mesh layout of the model: the fluid mesh is graded towards the free surface area.

2.3 Fluid solution

The fluid domain is governed by the Navier-Stokes equations for incompressible, isothermal, laminar flow, as expressed in Equation (7) and (8).

$$\nabla \cdot \boldsymbol{v} = 0 \tag{7}$$

$$\frac{\partial(\rho \boldsymbol{v})}{\partial t} + \nabla \cdot (\rho \boldsymbol{v}\boldsymbol{v}) = -\nabla P + \nabla \cdot (\boldsymbol{\tau}) + \rho g \tag{8}$$

where $\boldsymbol{v}$ is velocity vector, $P$ is pressure, $\rho$ is the density, $\boldsymbol{\tau} = \mu(\nabla \boldsymbol{v} + \nabla \boldsymbol{v}^T)$ is the viscous stress, $\mu$ is the dynamic viscosity and $g$ is gravitational acceleration set as 9.81 m/s².

The VOF method was used to capture the free surface between air and water. The VOF method introduces a passive scalar field $\alpha$, which denotes the fractional volume of a cell occupied by a specific phase. In this model, a value of $\alpha = 1$ corresponds to a cell full of water and a value of $\alpha = 0$ indicates a cell full of air. Thus, the free surface, which is a mix of these two phases, is formed by the cells with $0 < \alpha < 1$. A linear approximation was applied to satisfy the continuity of the free surface, as expressed in Equation (9) (Weller, 2002), and the evolution of free surface with time was solved by the advection equation of $\alpha$, as expressed in Equation (10) (Rusche, 2003). Furthermore, the local density and viscosity were determined according to Equation (11) and (12).

$$\boldsymbol{v} = \alpha \boldsymbol{v}_{water} + (1-\alpha)\boldsymbol{v}_{air} \tag{9}$$

$$\frac{\partial \alpha}{\partial t} + \nabla \cdot (\boldsymbol{v}\alpha) + \nabla \cdot [\boldsymbol{v}_r \alpha(1-\alpha)] = 0 \tag{10}$$

$$\rho = \alpha \rho_{water} + (1-\alpha)\rho_{air} \tag{11}$$

$$\mu = \alpha \mu_{water} + (1 - \alpha)\mu_{air} \tag{12}$$

where $v_{water}$ and $v_{air}$ are the velocities of the nearest water cell and air cell respectively and $v_r = v_{water} - v_{air}$ is the relative velocity between them (Vukčević, 2016). In this study, $\rho_{water}$ = 1000 $kg/m^3$; $\mu_{water}$ = 1 × $10^{-3}$ $N \cdot s/m^2$; $\rho_{air}$ = 1$kg/m^3$; $\mu_{air}$ = 1.48 × $10^{-5}$ $N \cdot s/m^2$.

2.4 Structural solution

The deformation of the ice sheet is governed by conservation of linear momentum, where the stress is given by the nonlinear St. Venant Kirchhoff hyperelastic law, as implemented by Tukovic et al. (2018) and Cardiff et al. (2018). The mathematical model in total Lagrangian form (reference configuration) may be written as:

$$\oint \rho_{ice} \frac{\partial}{\partial t}\left(\frac{\partial u}{\partial t}\right) dV = \oint \mathbf{n} \cdot (\Sigma \cdot \mathbf{F}^T) \, dS + \oint \rho_{ice} \mathbf{g} \, dV \tag{13}$$

where $\mathbf{u}$ is the displacement vector, and $u_x$ was prescribed at 0 at the middle point of the ice bottom to avoid the ice sheet from drifting. $\mathbf{F} = \mathbf{I} + (\nabla \mathbf{u})^T$ is the deformation gradient tensor, and $\mathbf{I}$ is the second-order identity tensor and $\Sigma$ is the second Piola-Kirchhoff stress tensor, which is related to the Cauchy stress tensor σ through Equation (14).

$$\sigma = \frac{1}{\det \mathbf{F}} \mathbf{F} \cdot \Sigma \cdot \mathbf{F}^T \tag{14}$$

The stress/strain relationship of the ice sheet is dictated as below,

$$\Sigma = 2G\mathbf{E} + \Lambda \, \text{tr}(\mathbf{E})\mathbf{I} \tag{15}$$

where the Green-Lagrange strain tensor is $\mathbf{E} = \frac{1}{2}[\nabla \mathbf{u} + (\nabla \mathbf{u})^T + \nabla \mathbf{u} \cdot (\nabla \mathbf{u})^T]$, and G and Λ are the Lamé's coefficients, related to the material properties of Young's modulus E and Poisson's ratio *v*, as:

$$G = \frac{E}{2(1+\nu)} \tag{16}$$

$$\Lambda = \frac{\nu E}{(1+\nu)(1-2\nu)} \tag{17}$$

2.5 Fluid-Structure Interaction

The partitioned FSI scheme of Tukovic et al. (2018) was applied to couple the fluid and solid solutions, where the fluid and structural equations were solved separately and coupled at the fluid-solid interface. A Dirichlet-Neumann coupling procedure is employed, where velocity and pressure are first calculated in the fluid sub-domain, and the force at the fluid side of the interface is applied as a boundary condition to the solid side of the interface; the displacement in the solid sub-domain is then solved and the velocity at the solid side of the interface is applied as a boundary condition to the fluid side of the interface. Iterations are performed over these steps until the interface kinematic and dynamic conditions are satisfied, as illustrated in Figure 3. The procedure is explained in detail as follows:

• To start a new time step loop, the structural displacement is first updated according to the results of the previous timestep. Then, to improve the convergence speed of the coupling procedure, the Aitken coupling scheme is employed, which introduces an Aitken Relaxation Factor (ARF), as:

$$\text{ARF}_{i+1} = \text{ARF}_i \times \left[1 - \frac{\sum(Res \cdot \Delta Res)}{\sum(\Delta Res \cdot \Delta Res)}\right] \tag{18}$$

Thus the ARF is updated according to the residual (Res), which is the difference between the structural interface displacement (SID) and the fluid interface displacement (FID), namely Res = SID − FID. Afterwards, the fluid mesh is adjusted with the updated ARF value, as:

$$\text{Fluid Mesh}_{i+1} = \text{Fluid Mesh}_i + \text{ARF} \times \text{Res} \tag{19}$$

The FID is extracted from the adjusted fluid mesh. Then its differential produces the velocity of the fluid interface and the mesh motion of the rest of the fluid mesh is obtained according to this interface velocity.

• Based on the moved mesh, the fluid solver calculates the velocity and pressure field. Then, the pressure and viscous force on the fluid interface can be obtained.

• The fluid load on the fluid interface is transferred to the solid interface.

• According to the load on the interface, the structure solver calculates the displacement of the solid.

• The SID can be extracted from the structural displacement and then compared with the FID to obtain a new residual, Res. The solver switches to the next time step when either the residual criteria is satisfied or the pre-defined maximum number of FSI outer iterations have been reached, otherwise it continues looping in the current time step.

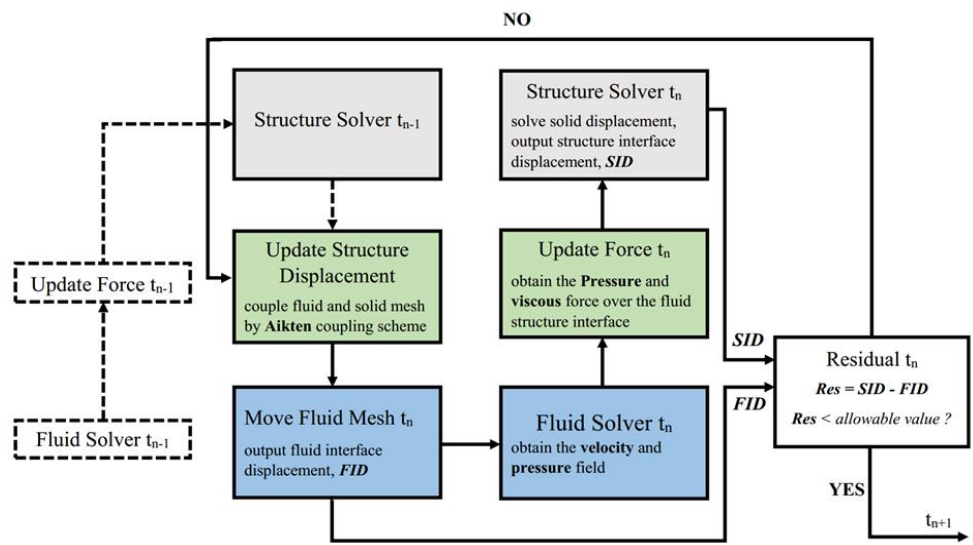

Figure 3: Flowchart of the FSI solution procedure.

## 3. Results and discussion

In this section, validation is presented to show the rationality and accuracy of the proposed numerical approach, following four steps: (a) ensure target wave field can be generated; (b) ensure wave-induced ice deformation can be accurately calculated; (c) without overwash, correctly predict the influence of the ice sheet on wave transmission and reflection; (d) including overwash, correctly simulate the wave-ice interaction. Once the numerical approach has been validated, extended investigations are presented to discuss the influence of overwash.

3.1 Mesh sensitivity tests

The FSI model was verified through two sensitivity tests for the fluid mesh and solid mesh respectively. Both aim to achieve accurate solutions with as few cells as possible, so as to minimise the computational costs. The first test was conducted to ensure the target wave field can be obtained without the presence of the ice sheet. For waves with large wavelengths and small amplitudes waves that are of interest, the wave field can be sensitive to the vertical mesh density at the free surface area. Therefore, the fluid cell number per wave height (M) was varied to test this influence. For the mesh density in the wavelength direction, 100 cells per wavelength was used for all simulations, with which mesh-independent results were produced. To obtain the wave field, the ice sheet was taken away and the free surface elevation was recorded using the two probes illustrated in Figure 1. With different M values, the recorded free surface elevation is presented in Figure 4, alongside the target value. It is shown that the target wave field can be obtained at both upstream and downstream locations of the ice sheet when M ≥ 15, which means a high quality of wave generation and propagation throughout the computational domain. Therefore, this study chose M = 15 for the fluid mesh density.

The second test was conducted to select an appropriate mesh density to capture the elastic deformation of the ice sheet. The solid mesh density was globally scaled, resulting in four meshes consisting of 200, 400, 800 and 1200 cells respectively. With the four meshes, the wave-induced deformation was measured at different locations of the ice sheet, and the measured vibration amplitudes are presented in Figure 5, alongside a comparison with the experimental data of Sree et al. (2017). It shows the ice sheet is undergoing significant deformation near the edges while the vibration amplitude is small in the middle area: this shows the ice sheet is bending due to the waves. The relative deviations between computational and experimental results were calculated by Equation (20) and shown in Table 2(a). The model predictions generally agree with the experimental results. Deviations between model and experimental results are small when the solid cell number is larger than 400, so the mesh set of 400 cells was selected for further simulations, corresponding to 100 cells per ice length and 4 cells per ice thickness.

$$\text{Relative deviation} = \frac{Com. - Exp.}{Exp.} \times 100\% \tag{20}$$

3.2 Validation

Validation of the model includes two parts. The computational results were first compared with the experimental data of Nelli et al. (2017) to assess the accuracy of the model to predict wave reflection and transmission against the ice sheet. In this part, edge barriers were attached to the ice sheet to avoid wave overwashing, as shown in Figure 6(a). In the second part, the edge barriers were removed to enable overwash as shown in Figure 6(b), and the mean depth of the overwash water was recorded and compared with the experimental data of Skene et al. (2015).

In accordance with the experiments of Nelli et al. (2017), simulations were carried out with a range of wave conditions, from short to long and from gently-sloping to storm-like, listed in Table 1 as Cases 1-9. Wave transmission and reflection are denoted as the wave energy measured by the two probes, for which both probes were placed far enough from the ice sheet to avoid the disturbance from breaking waves. To process data, the free surface elevation measured by both probes was analysed by a Fast Fourier Transform to obtain the wave energy spectrum, and the wave energy was the integration of the energy spectrum over the frequency domain:

$$S(f) = \frac{1}{2}\int_0^\infty \hat{a}^2 \, df \tag{21}$$

$$E = \int_0^\infty S(f) \, df \tag{22}$$

where $\hat{a}(f)$ is the wave amplitude component at each frequency $f$. The wave energy at the upstream/downstream probe location was recorded as $E_{front}$ and $E_{rear}$, where $E_{rear}$ denotes the transmitted waves while $E_{front}$ is a superposition of the ice-reflected waves and the incident waves. Thus, the reflection (R) and transmission (T) coefficients were calculated based on their ratio to the incident wave energy $E_{incident}$:

$$R = \frac{|E_{front} - E_{incident}|}{E_{incident}} \tag{23}$$

$$T = \frac{E_{rear}}{E_{incident}} \tag{24}$$

The comparison between computational and experimental results is shown in Figure 7 and Table 2(b), where good agreement can be seen for all the examined wavelength and wave height conditions. With increasing wavelength, T increases and R decreases, which means a better transmission appears with longer incident waves. Noting that overwash was avoided at this stage, R and T are insensitive to changes in the wave height, agreeing with the assumption of the linear theory (Meylan and Squire, 1994). Moreover, R + T approximately equals to one in these cases without overwash, which means energy dissipation is negligible in such a situation.

A further validation was conducted to assess the capability of the model to simulate overwash. According to the experiments of Skene et al. (2015), the depth of overwash water was measured at the middle of the ice surface, using a set of wave conditions listed in Table 1 as Cases 10-18. The comparison between computational and experimental mean water depth is shown in Figure 8 and Table 2(b). It can be seen that overwash depth from the model reveals a consistent trend with the experiments: it increases with increasing wave height and wavelength. In this case, overwash increases with longer waves, which is opposite to the trend observed in the interaction of waves with a small circular ice floe, where overwash was found to decrease with an increased wavelength (Yiew et al., 2016; Huang and Thomas, 2018). This difference indicates overwash is also dependant on the shape and dimension of the ice, relative to the wave. A notable deviation exists for the case of the largest wave height, i.e. Case 18, where the model overpredicts the overwash depth. This may be attributed to the turbulence occurring with such waves of very large height, which was observed in the experiments of Skene et al. (2015). Turbulence is a source of energy dissipation that can lead to a weaker overwash, which has not been included in the current work. Future work could consider finding an appropriate turbulence model to predict the complex overwash flow occurring during storm-like conditions.

3.3 Overwash

To investigate the influence of overwash on R and T. Cases 1-9 were simulated again but this time the edge barriers were taken away, thus overwash was enabled for the same wave conditions. The comparison was made between the results with and without overwash, as shown in Figure 9. It can be seen that both the R and T with overwash are smaller than those without overwash. For R, the results without overwash are slightly higher than these with overwash. This can be attributed to the edge barriers that provide a better reflection and also result from the energy dissipation caused by the wave breaking at the ice edge without barriers. For T, the results are shown to be significantly reduced by overwash. The difference is most pronounced when $\lambda/L = 1.26$, while it is less obvious when $\lambda/L = 1$ and $\lambda/L = 1.56$; this indicates the reduction effect on transmission is not positively or negatively correlated to the ice length. This finding agrees well with the experiments of Nelli et al. (2017), as well as the analytical solutions of Meylan and Squire (1994) where R and T were predicted to follow a periodically parabolic trend with non-dimensional ice sheet length.

With overwash, the ice sheet splits the incoming waves into the overwash water on top and hydroelastic-effect waves underneath. After passing over the ice sheet, the overwash water cannot reform waves so is not able to effectively transmit downstream. Therefore, this part of the incident waves is dissipated from transmission and reflection, resulting in $R + T < 1$. Experimental work reached the consensus that this part of energy brings about inaccuracies in current theoretical models where overwash is ignored (Toffoli et al., 2015; Nelli et al., 2017), which is illustrated in Figure 9 as the analytical solutions are similar to the computational results when overwash was avoided. Since the overwash water depth links with the incident wave height, the energy dissipation increases with increasing wave height. This does not follow the linear theory where a changed wave height is uninfluential, which means the wave-ice interaction including overwash is nonlinear and cannot be included in a linear model. Although Skene et al. (2015, 2018) developed nonlinear shallow water equations that can predict the depth of the overwash water, their method is based on a one-way coupling, i.e. no back coupling from overwash to predict the surrounding fluid domain. From this point of view, the present model of two-way coupling proves to provide more realistic results for overwash modelling.

## 4 Conclusions

A multiphase FSI code has been developed upon the work of Tukovic et al. (2018) and Cardiff et al, (2018), by which a successful attempt has been presented to simulate the hydroelastic response of sea ice. A full coupling has been achieved between the solutions of wave field and ice deformation, and it entirely uses the FVM method within the framework of OpenFOAM. Relevant numerical theories and approaches have been introduced in detail, and the applied code is publicly accessible (Huang, 2018). The model has shown good agreement with experiments and provided more realistic results than current theoretical methods. It can accurately predict the wave transmission and reflection over a floating ice sheet and effectively remedy the theoretical inaccuracies due to the energy dissipation from the overwash process.

Compared with the experimental method, a convenience of the developed model is that it is easy to change environmental variables, especially when considering the difficulty of manufacturing sea ice. Although the simulations presented are in model-scale for the purpose of validation, in principle there is no reason that the same approach cannot be applied at full-scale. In future work, the proposed code could be used to study wave-ice behaviours such as the viscoelastic attenuation (Sree et al., 2018) and the effect of varying ice dimensions, ice rheology and water depth on polar wave motions. Moreover, the present model can be easily extended to incorporate fixed/floating structures to provide valuable estimates for Arctic engineering purposes.

**Acknowledgements**

Thanks go to Professor Guoxiong Wu and Mr Bojan Igrec at University College London (UCL), for providing valuable discussions and helping establish the collaboration between the authors. The first and second authors are grateful to Lloyds Register Foundation, UCL Faculty of Engineering Science and China Scholarship Council, for funding their PhDs. The authors also appreciate Professor Decheng Wan and his research group at Shanghai Jiao Tong University, who hosted the 13[th] OpenFOAM Workshop. This work is part of a project that has received funding from the European Union's Horizon 2020 research and innovation programme under grant agreement No 723526 - SEDNA: Safe maritime operations under extreme conditions; the Arctic case.

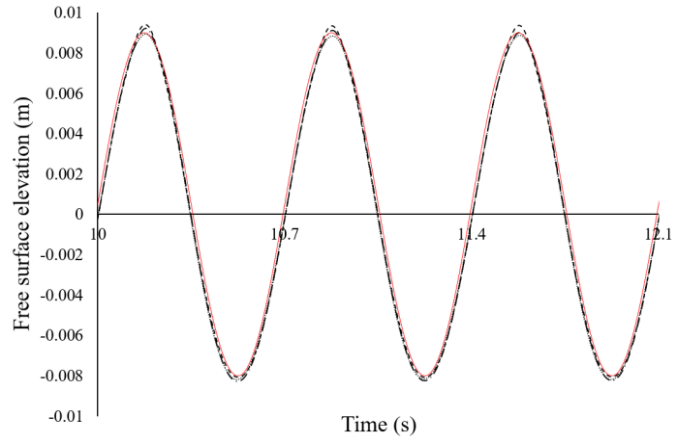

(a) Probe 1

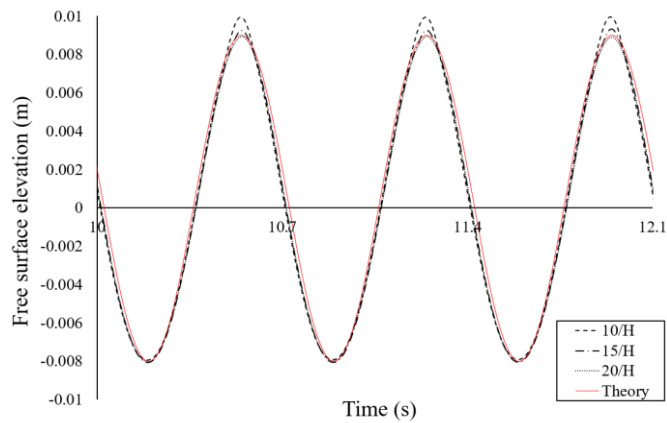

(b) Probe 2

Figure 4: Generated waves with different cell numbers per wave height. The target waves are of T = 0.7 s and H = 0.017 m.

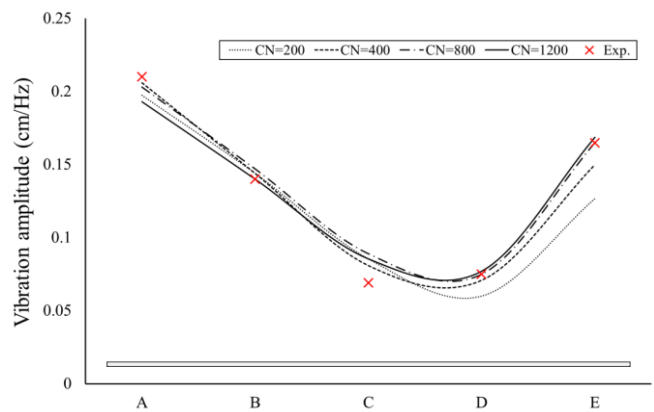

Figure 5: Vibration amplitude at different locations of the ice sheet, calculated with a range of solid cell number (CN). Experimental data of Sree et al. (2017) are also included. Applied wave and ice conditions are shown in Table 1: case 19. Points A – E are respectively 0.1m, 0.3m, 0.5m, 0.7m and 0.9m from the left edge of the ice sheet.

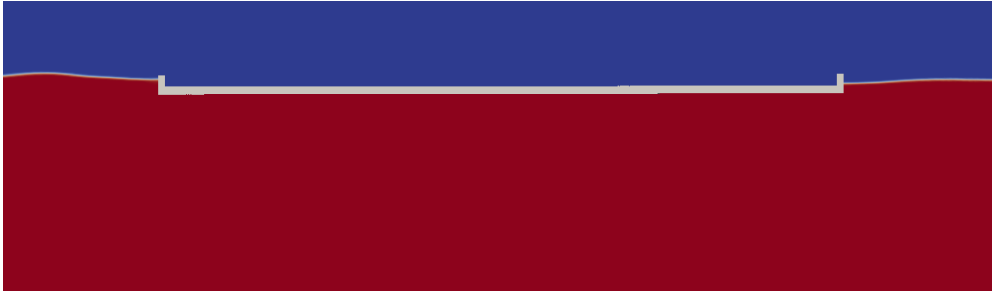

(a) Without overwash

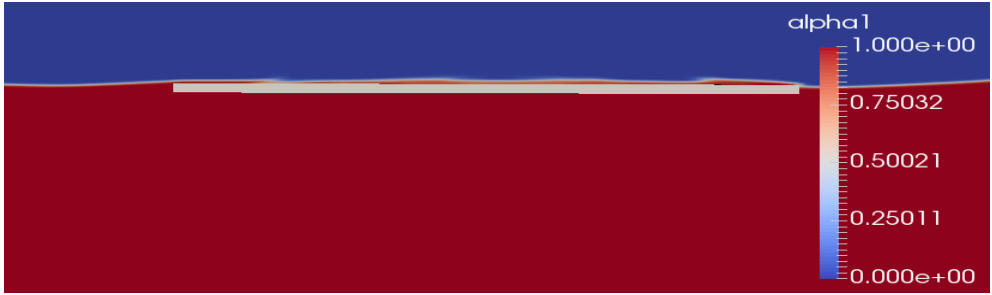

(b) With overwash

Figure 6: Examples of the wave-ice interaction in OpenFOAM.

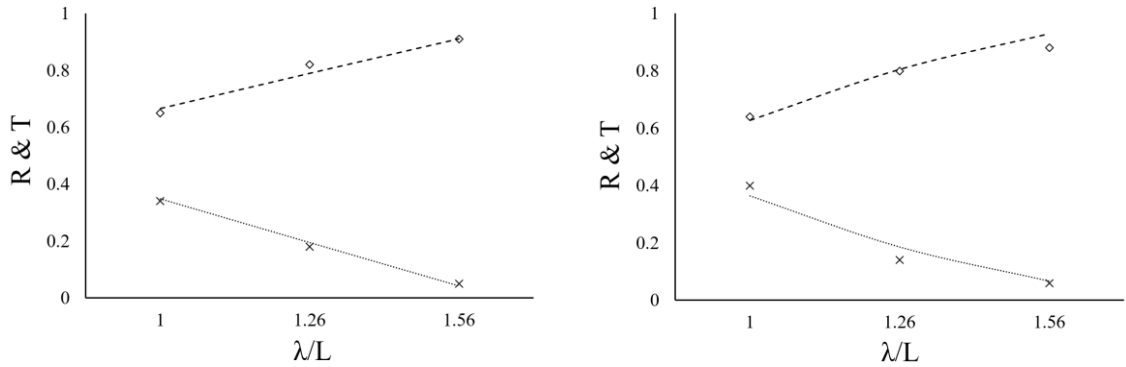

(a) ka = 0.08

(b) ka = 0.10

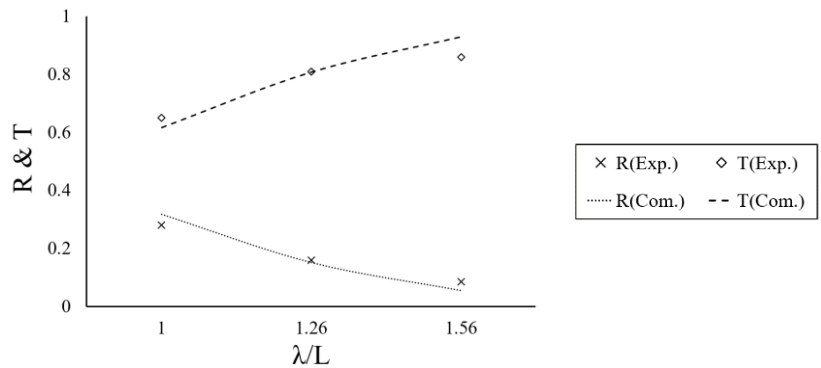

(c) ka = 0.12

Figure 7: Computational and experimental R & T (Nelli et al., 2017), obtained when overwash was avoided.

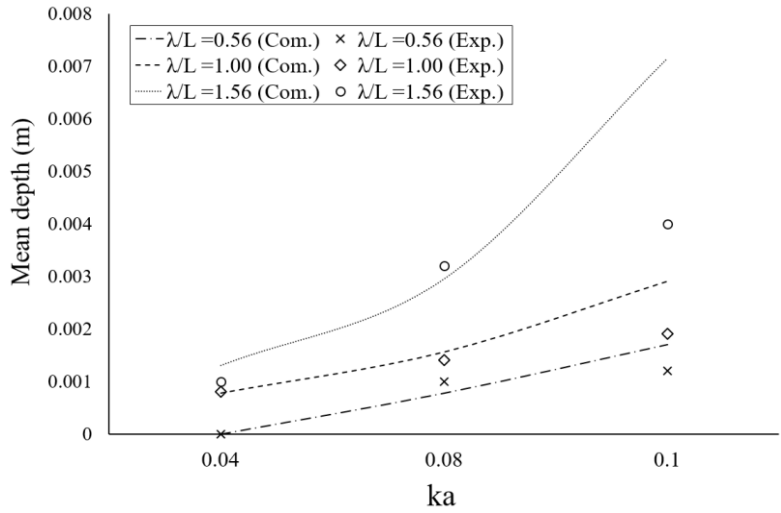

Figure 8: Computational and experimental (Skene et al., 2015) mean overwash depth.

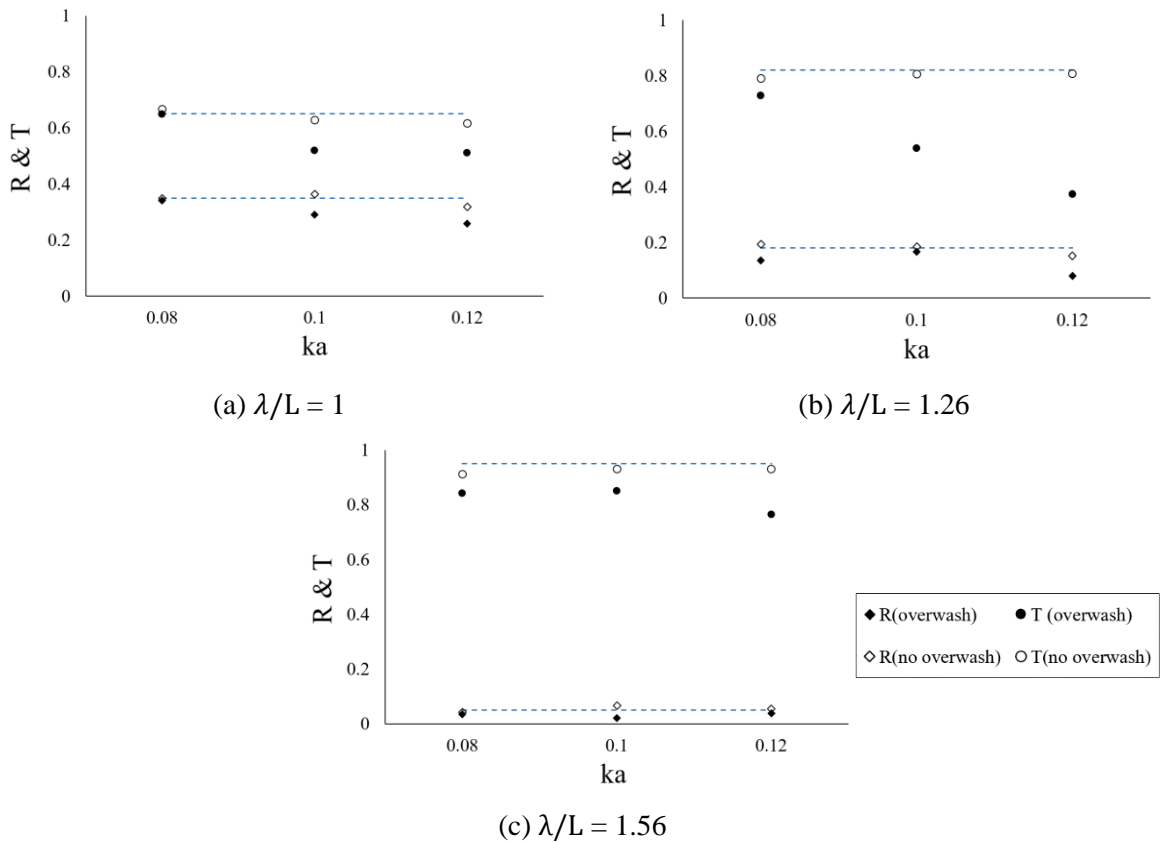

(a) $\lambda/L = 1$

(b) $\lambda/L = 1.26$

(c) $\lambda/L = 1.56$

Figure 9: R & T with and without overwash, alongside the analytical solutions of Nelli et al. (2017) (dash line).

Table 1: Parameters of the simulation cases

| Case | Wave condition | Ice condition | Water Depth |
|---|---|---|---|
| 1 | $\delta$ = 0.8 s, $\lambda$ = 1 m, H = 0.025 m (ka = 0.08) | L = 1 m, t = 0.01 m, $\rho_{ice}$ = 905 $kg/m^3$, E = 1.6 GPa, $\nu$ = 0.4, with & without edge barriers. | h = 0.9 m |
| 2 | $\delta$ = 0.8 s, $\lambda$ = 1 m, H = 0.032 m (ka = 0.10) | | |
| 3 | $\delta$ = 0.8 s, $\lambda$ = 1 m, H = 0.038 m (ka = 0.12) | | |
| 4 | $\delta$ = 0.9 s, $\lambda$ = 1.26 m, H = 0.032 m (ka = 0.08) | | |
| 5 | $\delta$ = 0.9 s, $\lambda$ = 1.26 m, H = 0.040 m (ka = 0.10) | | |
| 6 | $\delta$ = 0.9 s, $\lambda$ = 1.26 m, H = 0.048 m (ka = 0.12) | | |
| 7 | $\delta$ = 1.0 s, $\lambda$ = 1.56 m, H = 0.039 m (ka = 0.08) | | |
| 8 | $\delta$ = 1.0 s, $\lambda$ = 1.56 m, H = 0.049 m (ka = 0.10) | | |
| 9 | $\delta$ = 1.0 s, $\lambda$ = 1.56 m, H = 0.059 m (ka = 0.12) | | |
| 10 | $\delta$ = 0.6 s, $\lambda$ = 0.56 m, H = 0.007 m (ka = 0.04) | L = 1 m, t = 0.02 m, $\rho_{ice}$ = 905 $kg/m^3$, E = 1.6 GPa, $\nu$ = 0.4, without edge barriers. | h = 0.5 m |
| 11 | $\delta$ = 0.6 s, $\lambda$ = 0.56 m, H = 0.014 m (ka = 0.08) | | |
| 12 | $\delta$ = 0.6 s, $\lambda$ = 0.56 m, H = 0.018 m (ka = 0.10) | | |
| 13 | $\delta$ = 0.8 s, $\lambda$ = 1 m, H = 0.013 m (ka = 0.04) | | |
| 14 | $\delta$ = 0.8 s, $\lambda$ = 1 m, H = 0.026 m (ka = 0.08) | | |
| 15 | $\delta$ = 0.8 s, $\lambda$ = 1 m, H = 0.032 m (ka = 0.10) | | |
| 16 | $\delta$ = 1.0 s, $\lambda$ = 1.51 m, H = 0.019 m (ka = 0.04) | | |
| 17 | $\delta$ = 1.0 s, $\lambda$ = 1.51 m, H = 0.038 m (ka = 0.08) | | |
| 18 | $\delta$ = 1.0 s, $\lambda$ = 1.51 m, H = 0.048 m (ka = 0.10) | | |
| 19 | $\delta$ = 0.7 s, $\lambda$ = 0.755 m H = 0.017 m (ka = 0.07) | L = 1 m, t = 0.01 m, $\rho_{ice}$ = 910 $kg/m^3$, E = 870 MPa, $\nu$ = 0.3, with edge barriers. | h = 0.3 m |

Table 2 (a): Relative deviation of ice deformation with different solid mesh densities.

| | Point A | Point B | Point C | Point D | Point E |
|---|---|---|---|---|---|
| CN = 1200 | -7.0% | +0.1% | +24.4% | +2.6% | +2.2% |
| CN = 800 | -3.3% | +5.0% | +26.9% | +0.1% | +0.1% |
| CN = 400 | -1.9% | +2.8 | +21.6% | -5.0% | -9.0% |
| CN = 200 | -6.0% | +2.8% | +31.4% | -20.0% | -23.0% |

Table 2 (b): Relative deviations of R&T and mean overwash water depth.

|  | R | T |  | overwash depth |
|---|---|---|---|---|
| Case 1 | +2.3% | +2.4% | Case 10 | +0.1% |
| Case 2 | -9.1% | -1.8% | Case 11 | -17.2% |
| Case 3 | +13.3% | -5.1% | Case 12 | +36.7% |
| Case 4 | +8.0% | -3.7% | Case 13 | -3.0% |
| Case 5 | +22.5% | +0.6% | Case 14 | +11.4% |
| Case 6 | -5.4% | -0.2% | Case 15 | +44.1% |
| Case 7 | -15.1% | +1.7% | Case 16 | +23.0% |
| Case 8 | +11.4% | +5.6% | Case 17 | -7.7% |
| Case 9 | -25.6% | +8.1% | Case 18 | +77.8% |